\documentclass[aps,pre,reprint,superscriptaddress]{revtex4-1}
\pdfoutput=1

\usepackage{amsmath}
\usepackage{graphicx}
\usepackage{hyperref}

\usepackage{color}

\graphicspath{{figures/}}

\newcommand{\x}[1][]{\text{X}_{#1}}
\newcommand{\xp}[1][]{\text{X}^\prime_{#1}}
\newcommand{\y}[1][]{\text{Y}_{#1}}
\newcommand{\z}[1][]{\text{Z}_{#1}}

\newcommand{\f}[1][]{\mathrm{F}_{#1}}
\newcommand{\fs}[1][]{\mathrm{F}^*_{#1}}

\newcommand{\g}[1][]{\mathrm{G}_{#1}}
\newcommand{\gs}[1][]{\mathrm{G}^*_{#1}}

\newcommand{\m}[1][]{\mathrm{M}_{#1}}

\newcommand{\Le}{\text{L}}
\newcommand{\R}{\text{R}}

\newcommand{\A}{\text{A}}
\newcommand{\B}{\text{B}}
\newcommand{\C}{\text{C}}
\newcommand{\D}{\text{D}}
\newcommand{\E}{\text{E}}
\newcommand{\T}{\text{T}}

\newcommand{\rlh}{\rightleftharpoons}
\newcommand{\nn}{\nonumber}

\begin{document}

\author{Rory A. Brittain}
\email{roryabrittain@gmail.com}
\noaffiliation

\author{Nick S. Jones}
\affiliation{Department of Mathematics, Imperial College London, London, SW7 2AZ, UK}
\author{Thomas E. Ouldridge}
\affiliation{Centre for Synthetic Biology and Department of Bioengineering, Imperial College London, London, SW7 2AZ, UK}

\title{What would it take to build a thermodynamically reversible Universal Turing machine? Computational and thermodynamic constraints in a molecular design}

\begin{abstract}
We outline the construction of a molecular system that could, in principle, implement a thermodynamically reversible Universal Turing Machine (UTM). By proposing a concrete---albeit idealised---design and operational protocol, we reveal fundamental challenges that arise when attempting to implement arbitrary computations reversibly. Firstly, the requirements of thermodynamic reversibility inevitably lead to an intricate design. Secondly, thermodynamically reversible UTMs, unlike simpler devices, must also be logically reversible. Finally, implementing multiple distinct computations in parallel is necessary to take the cost of external control per computation to zero, but this approach is complicated the distinct halting times of different computations.     


\end{abstract}

\maketitle



Computational operations use physical substrates, and thus have physical consequences \cite{landauer1996physical}.  The equivalence of the Shannon and the non-equilibrium thermodynamic entropies \cite{esposito2011second} provides a concrete link between information-processing and thermodynamics; how the resultant thermodynamic costs might be minimised is an important open question.  Many analyses have focused on specific tasks, such as erasing and copying bits \cite{landauer1961irreversibility,szilard1929entropieverminderung}; recent studies have extended these ideas to the the processing of strings \cite{mandal2012work,boyd2016identifying,Stopnitzky2019Physical,brittain2019biochemical}. Others have focused instead on computation as an arbitrary process in which an input state is converted into an output. Within this framework, Bennett argued that thermodynamically reversible computers must use reversible logic, wherein the input state  can be inferred unambiguously from its output \cite{bennett1973logical}. However, it has since been shown that any input-output map---whether logically reversible or not---can be implemented in a thermodynamically reversible fashion \cite{owen2018number}, provided that the distribution of input states is known \cite{kolchinsky2017dependence}.

Turing machines are a model of arbitrary computation  \cite{turing1937computable}. Traditionally, the machine's memory is  a tape: a sequence of symbols from a finite alphabet. The tape is processed by a head with a finite state set that reads the symbol at a position on the tape. Then, based on a machine-dependent set of transition rules, the head writes a new symbol at that tape site, moves either left or right to an adjacent site, and the head state changes. Sufficiently complex machines can be `Turing complete': they can compute  any computable function. These machines are called universal Turing machines (UTM).

The thermodynamics of abstract UTMs has been explored in the context of algorithmic information theory \cite{kolchinsky2019thermodynamic}. However, it is also illuminating to consider actual physical designs of low thermodynamic-cost UTMs \cite{brittain2019biochemical,ouldridge2018power,kolchinsky2019thermodynamic}. By making even an idealised design explicit, one can avoid hidden violations of the second law, and reveal fundamental challenges in constructing a device.
 
 Molecular systems are promising substrates for computing \cite{bennett1982thermodynamics}; the similarity of tape-processing ribosomes and TMs is tantalising. One approach to molecular computation  uses macroscopic quantities of molecules for each computation, and either repeatedly intervening to direct the computational steps \cite{adleman1994molecular, beaver1995universal, rothemund1995dna,smith1995dna,currin2017computing} or allowing the solution to relax to equilibrium and perform a computation \cite{chen2013programmable,cardelli_chemical_2016}. In both cases, the bulk scale makes  the thermodynamic cost per computation large.
 
An alternative is to use a small number of computational molecules, coupled to large baths of ancillary fuel. Bennett sketched a UTM of this kind~\cite{bennett1973logical}, and more recently Qian, Soloveichick and Winfree \cite{qian2010efficient} conceived a specific DNA-based realisation of a stack machine, a close analogue of a UTM. In both designs, the overall free-energy change per step in the forward direction, $-\epsilon$, is constant during a computation. For $\epsilon>0$, the computation  proceeds irreversibly forwards with a finite entropy production (or thermodynamic cost) per step \cite{strasberg2015thermodynamics}. For $\epsilon \rightarrow 0$, the computation undergoes unbiased diffusion, and the output computational state will not dominate the ensemble even at infinite time. Moreover, the total entropy production  is positive even in this unbiased limit, due to the irreversible spreading of the system's probability distribution over its computational states  \cite{strasberg2015thermodynamics}.

We outline a thermodynamically reversible, sequentially-operated molecular UTM in which the underlying computational logic is implemented by an explicit model of a surface-localised chemical reaction network. We first outline the general design principles, before describing the device in more detail. We then  illustrate key challenges that arise when implementing thermodynamically reversible UTMs, as opposed to simpler operations. Firstly, even a minimal design is intricate, emphasising how challenging it would be to actually construct a thermodynamically reversible, sequentially-operated molecular UTM. Secondly, thermodynamically reversible, sequentially-operated UTMs---unlike simpler devices---must be logically reversible. Finally, running distinct computations in parallel to minimise the costs of control is complicated by the distinct halting times of different computations.     

{\em Control protocols for thermodynamically reversible UTMs}. A reversible UTM must be subject to time-dependent external control, since any free evolution of a  thermodynamic system is necessarily irreversible \cite{machta2015dissipation,ouldridge2018importance}. However, systematic evolution of the control variable itself implies entropy generation \cite{machta2015dissipation}. In traditional thermodynamics, controls are applied to large systems: costs associated with evolving the control variable itself, and implementing any logic therein, can be made small relative to changes in the system \cite{machta2015dissipation}. The same is not true for controlling a single computation. The same control must therefore be applied to multiple UTMs in parallel  \cite{brittain2019biochemical}, so that any thermodynamic costs of applying a control protocol are negligible when divided by the number of computations. Controls cannot, therefore, include feedback on the state of individual computations, since the costs arising cannot be spread over multiple UTMs.


{\em Ideal molecular reactions provide an explicit model system from which to construct a thermodynamically reversible UTM.}  We assume that it is possible to design chemical reaction networks with reactions of the form

%

\begin{equation}
    \sum_{\{ i \}}\x[i] +\sum_{\{j \}}\y[j]+\fs\rlh    \sum_{\{i\}} \xp[i]+ \sum_{\{j\}} \y[j] +\f.
    \label{eq:reactionwithcatalyst}
\end{equation}
Here, $\x[\{i\}]$ is a set of substrate species converted into products $\xp[\{i\}]$ by the action of catalysts $\y[\{j\}]$. Each reaction is  driven by the turnover of a fuel molecule $\fs\rlh \f$. While such reactions are not elementary, compound catalytic action occurs naturally in, for example, transcription and translation, and can be engineered using DNA nanotechnology \cite{qian2010efficient,chen2013programmable,srinivas2017enzyme}. These reactions need not approximate mass-action kinetics \cite{plesa2018stochastic}, but the overall stoichiometry must be tightly observed, with the reactions only occurring if all catalysts and fuel are present. 

\begin{figure}[t]
	\def\svgwidth{\linewidth}
	\input{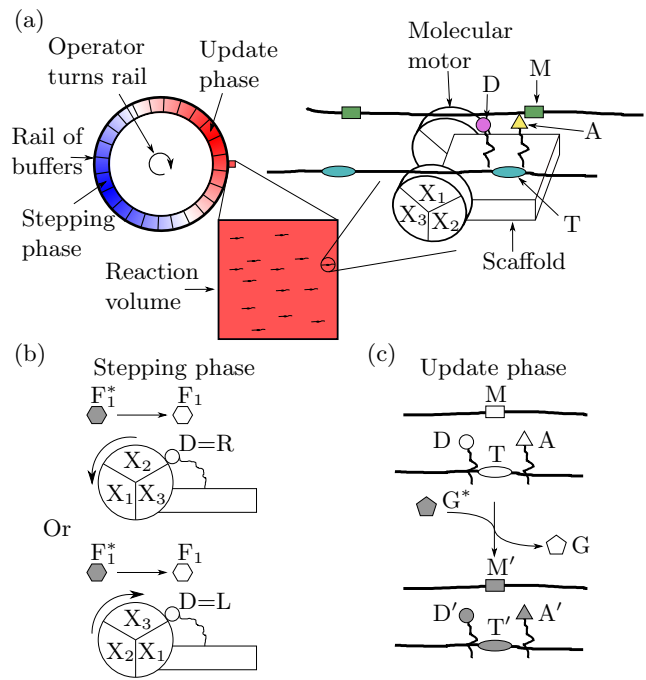}
    \caption{\emph{Overview of a molecular UTM.} (a) A reaction volume containing many UTMs. A protocol of fuel concentrations is applied by connecting the volume to a series of large, fuel-containing baths. The baths are arranged in a circle so the same protocol can be applied repeatedly. Individual UTMs act on two tape polymers with evenly spaced molecules (M,T) to represent their state. The state of the UTM head is encoded in the state of molecules (A,D) attached to a scaffold such as DNA origami. The head is attached to the tapes by molecular motors that move the head from one symbol to the next. (b) Stepping is powered by fuel molecules F$_i$. The directional molecule D on the head catalyses stepping in the appropriate direction. (c) During the update phase, states of the head (A), tape molecules (M,T), and the directional molecule (D) are updated, powered by a fuel molecule $G$.}
    \label{fig:overview}
\end{figure}

 We consider many UTMs in a single reaction volume (Fig.~\ref{fig:overview}\,(a)), all coupled to the same 
 large baths  of fuel (the external control) via a semi-permeable membrane. By varying the concentration of these baths \cite{schmiedl2007stochastic,rao2016nonequilibrium,ouldridge2017fundamental}, reactions of the form of Eq.~\ref{eq:reactionwithcatalyst}, involving the UTM's species, can be quasistatically driven in either direction, or turned on and off \cite{brittain2019biochemical}. Our design of a molecular UTM is sequential: it has clocked control cycles involving a single update of the head, tapes and the head position. All cycles are driven by the same sequence of baths, which can be arranged in a circle as shown in Fig. \ref{fig:overview}(a). One turn of the circle corresponds to one step of each UTM. 

This setup has several conceptual advantages \cite{ouldridge2018power}. Firstly, providing the control is simple; we  cycle a series of buffers past a reaction volume, without adjusting to the system's response; the control is mechanical but the computation is chemical.
Secondly, the baths both control the reaction and act as a reservoir of chemical work; there is no ambiguity about how work is stored and transferred. Thirdly, the reaction in Eq.~\ref{eq:reactionwithcatalyst} allows the $\y[\{j\}]$ and $\x[\{i\}]$ species to influence the evolution of the $\x[\{i\}]$ species. However, the reaction only occurs if the fuel is present; quasistatic re-wiring can be performed by changing which fuel molecules are present \cite{ouldridge2018power, brittain2019biochemical}.

\begin{figure}[t]
	\def\svgwidth{\linewidth}
	\input{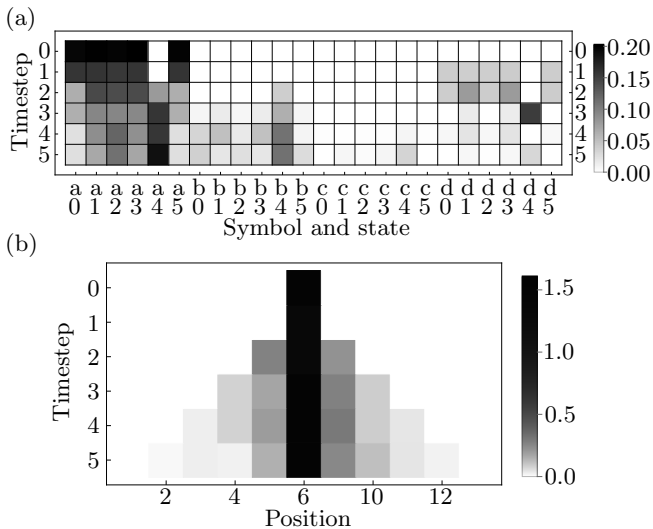}
    \caption{\emph{UTM operation generates both a complex distribution of head and tape states at each step, and correlations between states on the tape.} We consider a 4-state, 6-symbol machine (Appendix~\ref{ap:Rogozhin}) (a) Evolution of the probability distribution of the head and active tape site at the stare of each update phase. The head is initially in state A and positioned in the middle of 5 non blank symbols in an infinite tape of blank symbols. The machine is run for 5 steps with the states of the 5 non-blank symbols randomly chosen for each of 100,000 iterations. (b) The mutual information between the symbol at position 6, the initial position of the head, and the symbols at the neighbouring positions at the start of each update. The head is initially in state A and at position 6 in the middle of 13 non-blank symbols in an infinite tape of blank symbols. The machine is run for 5 steps with the states of the 13 symbols randomly chosen for each of 100,000 iterations.}
    \label{fig:distributionandinformation}
\end{figure}



\emph{Design and operation of a molecular UTM.} The head of a UTM in our design consists of two or more molecular species localised to a scaffold surface (Fig.~\ref{fig:overview}(a)).
The head interacts with two tapes, a main tape and a memory tape, represented by molecules attached to polymers at regular intervals. The head is attached to the polymers by molecular motors. At any one time, the head molecules can interact with each other and at most one pair of sites on the tapes. Since we are concerned only with the thermodynamics of converting inputs to outputs, we assume that the machines and tapes are provided in their starting configurations.

The UTM cycle has distinct stepping and update phases. During the stepping phase (Fig.~\ref{fig:overview}(b)),
The head must step selectively in either direction along the main tape
without relying on the discretion of an operator. This directionality is implemented through a `direction molecule' $\D$, which catalyses the transition of the motor between its configurations when the appropriate fuels are present. The direction molecule can either be in the left, $\Le$, or the right, $\R$, state, specifying the direction of the next step. The direction molecule also has a halting state $H$ that does not catalyse stepping on the main tape.

If the stepping of the motor involves at least three different substates, $\x[1]$, $\x[2]$ and $\x[3]$ (Fig.~\ref{fig:overview}(b)), fuels can be added that drive $\x[1] \overset{\Le}\rightarrow \x[2] $ and $\x[1] \overset{\R}\rightarrow \x[3] $;  $\x[2] \overset{\Le}\rightarrow \x[3] $ and $\x[3] \overset{\R}\rightarrow \x[2] $; and $\x[3] \overset{\Le}\rightarrow \x[1] $ and $\x[2] \overset{\R}\rightarrow \x[1]$. If these fuels are introduced and removed quasistatically and sequentially, stepping will be deterministic and thermodynamically reversible; one step is shown in Fig.~\ref{fig:overview}(b) and further details are given in  Appendix~\ref{ap:stepping}. The motor attached to the memory tape always takes a single step in the same direction during the stepping phase. 

The head state, the symbols on the tapes and direction molecule are updated during the update phase via reactions of the form of Eq.~\ref{eq:reactionwithcatalyst} (Fig.~\ref{fig:overview}(c)). The details depend on the specific set of rules chosen for the UTM, and how the states of the head are encoded in the molecular species. In the conceptually simplest case, a single molecule $\A$ represents the head state, along with the direction molecule. In this case, each update rule of a UTM can be expressed through reactions of the form 
\begin{align}
	\A+\Le+\T+\m[0]+\gs&\rlh \A^\prime+\D^\prime + \T^\prime+\m[2i]+\g\nn, \\
	\A+\R+\T+\m[0]+\gs&\rlh \A^\prime+\D^\prime + \T^\prime+\m[2i-1]+\g,
	\label{eq:update_head}
\end{align}
where $\m[0]$ is the default initial state of the memory tape and $i$ numbers each transition rule; the memory tape therefore makes a record of the transition taken. A and A$^\prime$, T and T$^\prime$, and L/R and D$^\prime$ represent the initial and final states of the head, the main tape and the direction molecule, respectively. The protocol involves simply modulating fuel concentrations in the following manner
\begin{equation}
	\begin{matrix}
		[\g]: & 0 & \rightarrow & g & \rightarrow & g & \rightarrow & 0 & \rightarrow & 0 \\
		[\gs]: & 0 & \rightarrow & 0 & \rightarrow & \gg g & \rightarrow & \gg g & \rightarrow & 0,
	\end{matrix}
\end{equation}
where $g$ is a large concentration. This protocol drives the reactions in Eq.~\ref{eq:update_head} reversibly from left to right, then switches them off before stepping.

In this set-up, each transition rule of the UTM needs two reactions. Each rule also requires two different memory states, in addition to $\m[0]$. The directional molecule has three states. The number of head molecule and tape states follow from the specific UTM. It is possible to use fewer memory species at the expense of increased conceptual complexity; the reactions and species required for some example machines are shown in Appendix~\ref{app:examples}. It is apparent that even the simplest design is sophisticated, with molecular species involved in several highly specific reactions. The need to perform precisely the right series of sequential reactions, and couple those reactions tightly to motion along a polymer, makes this complexity unavoidable. Without even considering the challenges of applying quasistatic manipulation of the baths, this complexity far exceeds what is currently implementable in practice. This complexity is easily overlooked in models that do not represent every stage explicitly \cite{bennett1982thermodynamics}.



{\em An ensemble of thermodynamically reversible UTMs requires an ensemble of distinct input programs.}
Fig.~\ref{fig:overview} shows an ensemble of UTMs operating in parallel. If the task was to erase or extract work from inputs, then it would be reasonable to  act in parallel on identical inputs. For each additional system, something more is achieved. However, if the goal is to answer a computational question, then acting in parallel on identical inputs does not reduce the cost per answer. Spreading the costs of external control over many computations requires parallel operation on multiple different inputs.

The first consequence of performing distinct computations in parallel is that a record of each computation must be unambiguously associated with each tape. If not, there would be no way to match the answers to questions asked. This record could be physically attached to the tape. Note that the record is a specification (eg. ``divide 135 by 5 and output the answer in a specific way"), so need not be a copy of the program itself.

{\em  Sequentially-operated, thermodynamically-reversible UTMs must have logically-reversible update rules.}
The memory tape makes the computation locally logically reversible: at each stage the input state of head and tape molecules can be unambiguously inferred from their final values. Without the memory, multiple  inputs would be mapped to the same output by the reactions of Eq.~\ref{eq:update_head}. This mixing  defines logical irreversibility. 

Consider  $\A+\Le+\T$ and $\A^\prime+\Le+\T$ being converted to the same output. At some instant, systems that started in $\A+\Le+\T$ will transition to the states occupied by systems that started in $\A^\prime+\Le+\T$, and vice versa. At this time, if a net flow of trajectories occurs in either direction, the process is thermodynamically irreversible \cite{ouldridge2018importance}. For logical states to mix reversibly, they must be offset in free energy to avoid net transitions from the more populated state to to the less populated one.
As a result,  a thermodynamically reversible protocol for a logically irreversible input-output map must be tuned to the distribution of input states \cite{kolchinsky2017dependence}. 

Tuning to the input distribution is far more problematic for
  sequentially-operated UTMs than
for simpler operations like erasing or copying of bits. Firstly, sequential UTMs implement complex computations through a series of simpler steps. To be thermodynamically reversible, a protocol would need to be optimised to the initial distribution of the head and tape {\em at each step} \cite{wolpert2015extending}. As illustrated in Fig.~\ref{fig:distributionandinformation}\,(a), this initial distribution has complex behaviour even for a simple UTM; re-applying the same cyclic fuel protocol, as in Fig.~\ref{fig:overview}\,(a), would be impossible. Instead, we'd need to encode  {\em a priori} knowledge of how the UTM would process a distribution of input tapes into a long, non-cyclic array of fuel baths. We'd effectively need to know the output of the relevant computations already just to build the UTM.
 
 Secondly, although the update rules of  UTMs are local, statistical information will exist between the locally active region of the UTM and other sites in the tape, as well as the record of the computation. As demonstrated in {Fig.~\ref{fig:distributionandinformation}}\,(b),  UTMs generate this information even if it does not initially exist. A local update of a subsystem when information exists between that subsystem and the rest of the system typically causes ``modularity dissipation" \cite{boyd2018thermodynamics,wolpert2020minimum}, wherein the non-equilibrium free energy stored in this information is wasted irreversibly. To avoid modularity dissipation, either: (a) no information with the rest of the system must be lost; or (b) the free energy stored in this information must be extracted. If the local update is logically reversible, (a) is trivially satisfied since the update simply permutes the occupancy of states. For logically irreversible updates, the mixing of distinct inputs will tend to reduce information with the rest of the system. To compensate, the rest of the system would have to be involved in the update---not to set the output, but to influence the protocol experienced by the active subsystem and thereby recover the stored free energy. For the the design in Fig.~\ref{fig:overview}, the rest of the tape would have to catalytically couple the head and active site to well-tuned fuel baths for information to be exploited. Doing so  would  be even more challenging---both mechanistically and computationally---than building a logically reversible device.

{\em The interplay of halting and thermodynamics.} To compute, a UTM must halt. In simpler contexts, when the same operation can be performed in parallel on multiple identical inputs, all operations will halt simultaneously. For UTMs performing arbitrary and distinct computations in parallel, however, halting times may be widely distributed; individual  halting times may not be known \emph{a priori}, and some computations may not halt at all.

This additional complexity raises three challenges. The first is that halted UTMs must stay in a valid halted state as the external protocol continues to be applied. This feature is designed into our molecular UTMs: the head and main tape do not evolve once the halt state is reached. The stepping is halted on the main tape but continues on the memory tape. Secondly, the control cycle must eventually stop, and it cannot simply intuit when all computations are halted. If an upper limit on the time of the computations in question is known, or if the operator is prepared to accept that some computations may not have halted or that unnecessary steps may be taken, the protocol can be stopped after a fixed number of steps with a cost that scales sub-linearly with the number of computations \cite{machta2015dissipation}.

Alternatively, although measurement and feedback on individual UTMs would violate the parallel nature of the protocol, the minimal {\em thermodynamic cost} of answering and acting upon the binary question of whether {\em all} machines have halted scales sub-linearly with the number of UTMs. However, the {\em mechanistic} challenges are high. For example, a single molecule  E that can undergo fuel-powered conversion catalysed by L or R, $\E \overset{\Le}\rightarrow \E^\prime$ or $\E \overset{\R}\rightarrow \E^\prime$, would reach the state $\E^\prime$ via a single reaction if and only if at least one of the UTMs had not halted. However, the operator would have to wait for a single probe to interact with all UTMs.

Thirdly, and most deeply, as the faster computations  halt, the protocol is applied to fewer active UTMs. The strategy of spreading the control cost over many computations is then questionable. In our case, control cost is the cost of ensuring that the baths move systematically past the reaction volume for the desired number of cycles; it is separate from the efficiency of the chemical computation. Machta argued that each unidirectional cycle of a control parameter has a minimal entropy generation associated with it \cite{machta2015dissipation}; we therefore assume that the cost of external control $C$ is linear in the number of protocol cycles $t$ implemented, $C(M,t) = \alpha(M) t + C_0(M)$, with $C_0$ representing initiation and termination costs and $M$ the number of distinct computations under control, a proxy for the system size. We allow for a dependence of $C$ on $M$: Machta found $\alpha(M) \sim M^{0.5}$ \cite{machta2015dissipation}.

If all $M$ computations in a set $\mathcal{S}_M$ halt in a time $t_M$, the minimal control cost per computation is $C(\mathcal{S}_M)/M = \alpha(M) t_M/M + C_0(M)/M$.
If the $M$ computations are drawn from a set $\mathcal{S} \supset \mathcal{S}_M$ of interest, we can ask whether it is always possible to reduce $C(\mathcal{S}_M)/M$ by using a subset with a larger  $M=|\mathcal{S}_M|$.
Assuming $C_0(M)$ scales sub-linearly with $M$ \cite{machta2015dissipation}, $C_0(M)/M \rightarrow 0$ as $M \rightarrow \infty$. However, as we increase $M,$ newly-chosen inputs may have a longer halting time and hence $t_M$ will grow. If the halting times of computational problems of interest are sparsely distributed, $\alpha(M) t_M/M$ may grow with $M$, implying that $C(\mathcal{S}_M)/M$ cannot be made negligible. 

For $\alpha(M) \sim M^{0.5}$, we require that subsets $\mathcal{S}_M$ can be found such that $t_M$ grows more slowly than $M^{0.5}$. Consider an algorithm with a halting time that is linear in the length $N$ of the input. If there are $N!$ problems of interest of length $ \leq N$, sets $\mathcal{S}_M$ can be found satisfying $t_M \sim M/{M}!$ and $C(\mathcal{S}_M)/M$  can easily be made negligible. If, by contrast, there is only one problem of interest at each length,  $t_M \sim M$ and $C(\mathcal{S}_M)/M$  grows with $M$.

Requiring that $C(\mathcal{S}_m)/M\rightarrow 0$ implies that all computations, regardless of their underlying difficulty, are equally valuable. An alternative would be to divide the cost per computation by a measure of the time complexity of the computations performed. As outlined in Appendix~\ref{ap:cost_per_time}, the resulting constraint is easier to satisfy but sets of computations $\mathcal{S}$ still exist for which the normalised cost of control grows with $M$ for $\mathcal{S}_M \subset \mathcal{S}$.


We have shown that the construction of a sequentially-operated, thermodynamically reversible molecular UTM has challenges that arise from the very purpose of a UTM: to perform complex computations. To circumvent some of these challenges, one could consider a machine that operates in a single control step. For example, the stack machine of Qian \emph{et al.} \cite{qian2010efficient} could migrate from the input to the output state if the driving from ancillary fuel molecules was slowly increased over time, avoiding the issue of choosing when to stop. However, doing so would raise a new problem: the later states would rapidly transition from being exponentially suppressed to exponentially favoured relative to the early states as the driving force was varied. For calculations of unknown (and arbitrary) length, it would be extremely challenging to implement a protocol that manages this change reversibly. In fact, in cases where the halting state is never reached, the computation would undergo an uncontrolled and irreversible growth as the driving force passed through zero. 


\bibliographystyle{mystyle}
\bibliography{bib}

\clearpage

\onecolumngrid

\appendix
\section{Stepping}
\label{ap:stepping}
We assume that the head changing position is an elementary transition that can happen in a chemical reaction: we assume that the molecular motor can rotate a third of a step while simultaneously switching which segment of the motor is bound to the polymer in a single step without the polymer and motor becoming unbound at any point. This is a strong simplifying assumption so a version of stepping without this assumption is given in Appendix \ref{app:alternativemolecularmotor}.

\begin{figure}
	\centering
	\includegraphics[width=\linewidth]{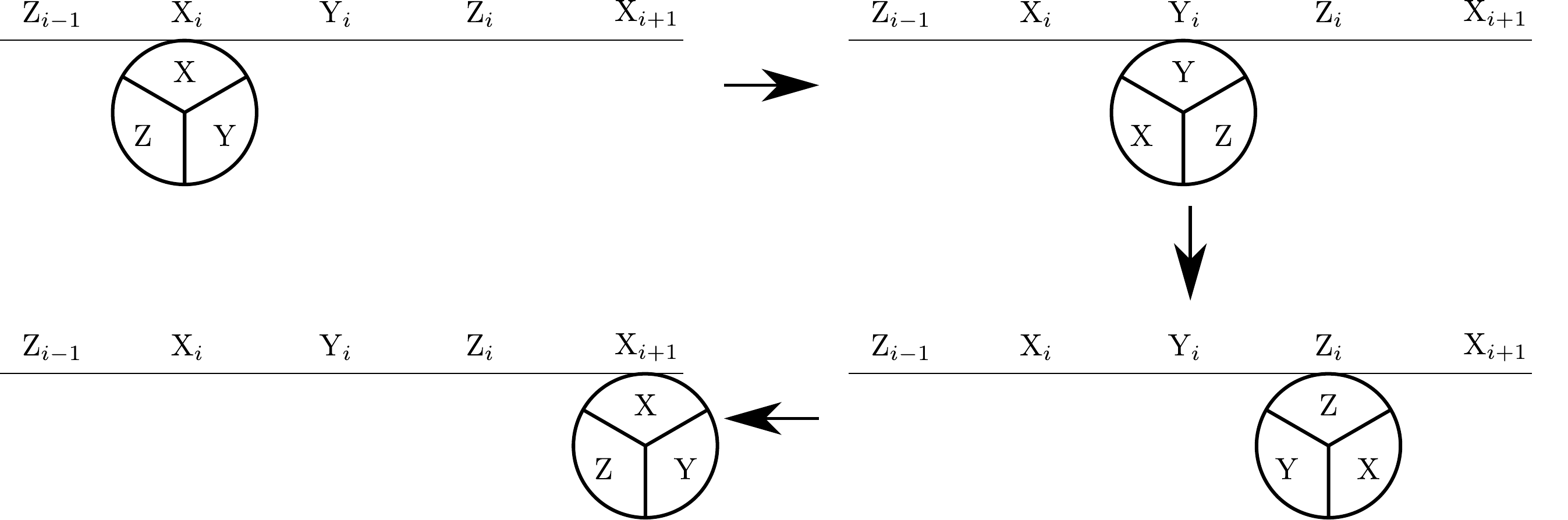}
	\caption{The stepping of the molecular motor.}
	\label{fig:stepping}
\end{figure}

The tape has a repeating structure of sites where the molecular motor can bind. There are three different types of site: X, Y and Z that can, respectively, bind to the X, Y and Z regions of the molecular motor. This is shown in figure \ref{fig:stepping}. The $\x$ sites are the positions on the tape where the symbol molecule on the tape lines up with the tethered head molecules (these are the only positions that correspond to positions in the abstract view of the machine) and the $\y$ and $\z$ are intermediate positions to which the head can attach. 

To move right the reactions

\begin{align}
	\fs[1]+\R+\x[i]&\rlh\R+\y[i]+\f[1],\nn\\
	\fs[2]+\R+\y[i]&\rlh\R+\z[i]+\f[2],\nn\\
	\fs[3]+\R+\z[i]&\rlh\R+\x[i+1]+\f[3],
	\label{eq:righ}
\end{align}
are used and to move left the reactions
\begin{align}
	\fs[1]+\Le+\x[i]&\rlh\Le+\z[i-1]+\f[1],\nn\\
	\fs[2]+\Le+\z[i-1]&\rlh\Le+\y[i-1]+\f[2],\nn\\
	\fs[3]+\Le+\y[i-1]&\rlh\Le+\x[i-1]+\f[3],
	\label{eq:left}
\end{align}
are used. If the head contains the direction molecule in the $\R$ state then the head can be quasistatically moved from $\x[i]$ to $\y[i]$ to $\z[i]$ to $\x[i+1]$ and if the direction molecule is in the $\Le$ state then the head can be quasistatically moved from $\x[i]$ to $\z[i-1]$ to $\y[i]$ to $\x[i-1]$. This is done using the protocol for three separate bit flips starting from a definite state.

For the left motion we could, alternatively, use three different pairs of fuel molecules, $\f[4]$, $\f[5]$ and $\f[6]$, from the right motion. This could make the mechanism easier to understand because the mechanisms for the two directions are independent. We adopt the approach in equations \ref{eq:left} and \ref{eq:righ} as reusing the same species reduces the total number of fuel molecule required.

\begin{figure}
	\centering
	\includegraphics[width=\linewidth]{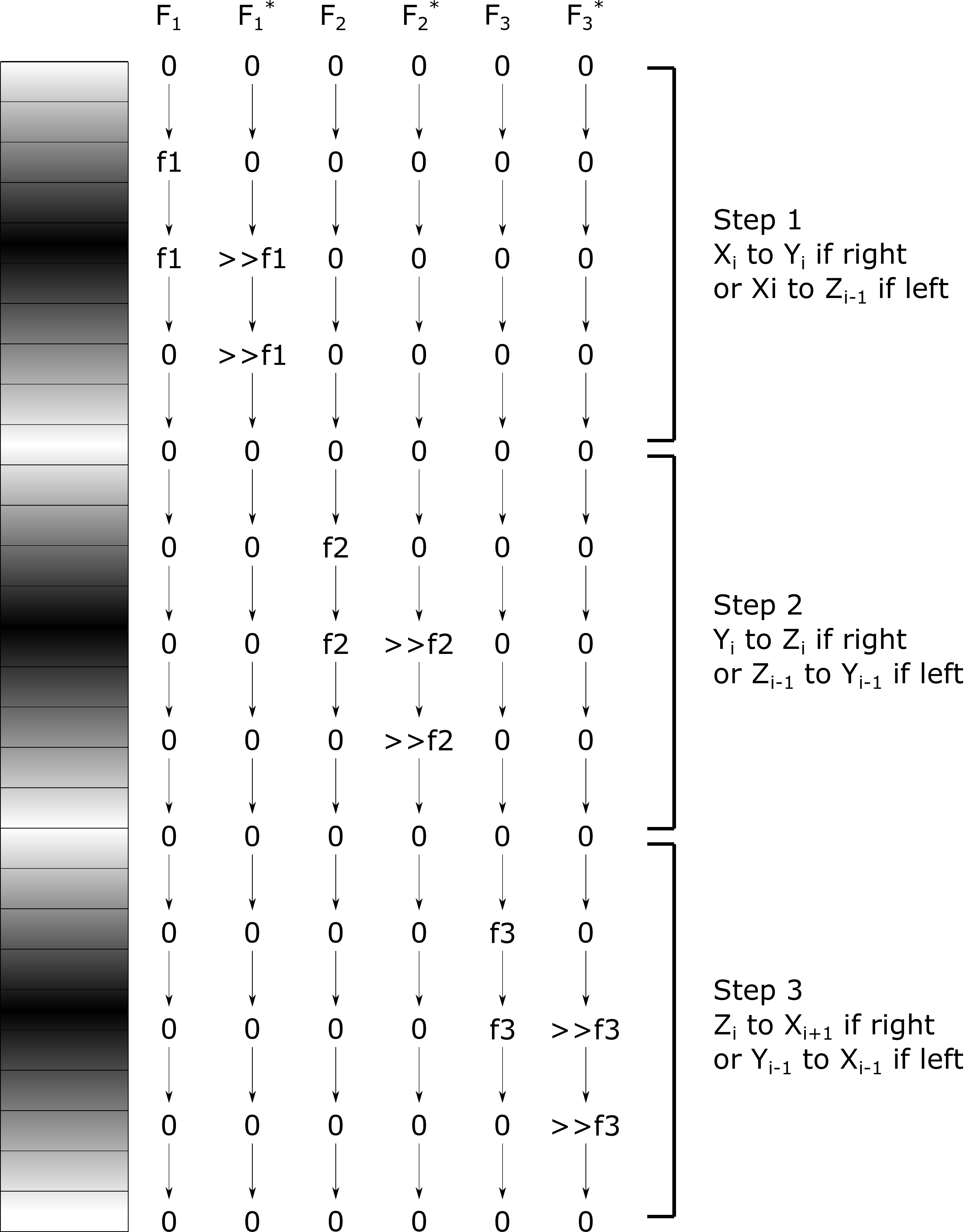}
	\caption{The protocol of concentrations of the fuel molecules to implement one step, which is left or right depending on the state of the direction molecule. While apparently complicated in fact the same approach is repeated three times. f1, f2 and f3 need only be a large concentration and in fact the same label could have been used in all cases.}
	\label{fig:steppingprotocol}
\end{figure}

The protocol of concentrations for the fuel molecules is shown in figure \ref{fig:steppingprotocol}. The work extracted from the buffers is the change in free energy of the head when it is attached to the different positions. We assume that the free energy is the same when the head is attached to any of the $\x[i]$. The free energy when attached to either of the intermediate states could be different but that is unimportant since the work done to move from $\x[i]$ to $\x[i+1]$ or $\x[i-1]$ is zero.

\section{Alternative Molecular motor}
\label{app:alternativemolecularmotor}
\begin{figure}
	\centering
	\includegraphics[width=\linewidth]{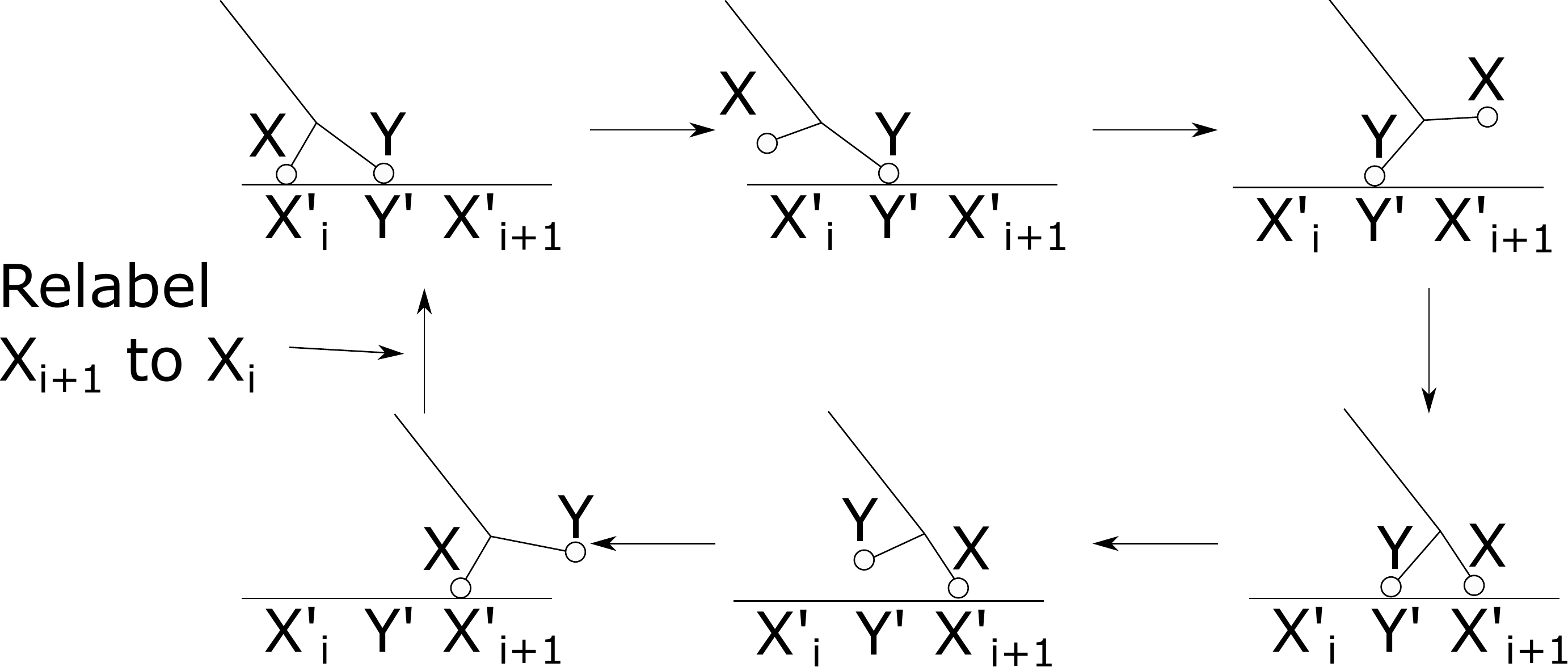}
	\caption{Overview of more realistic molecular motor. The arrows represent the direction of the quasistatic change when taking a rightwards step. Of course, each step is microscopically reversible and when moving left the arrows are all in the opposite directions. The X$^\prime$ is meant to show that that is a place where the X foot can bind.}
	\label{fig:walker}
\end{figure}

It is more realistic if the motor can detach from the tape. Consider an actin/myosin style set-up where the motor has two feet. An overview is shown in figure \ref{fig:walker}.

\begin{figure}[h!]
	\centering
	\includegraphics[width=0.55\linewidth]{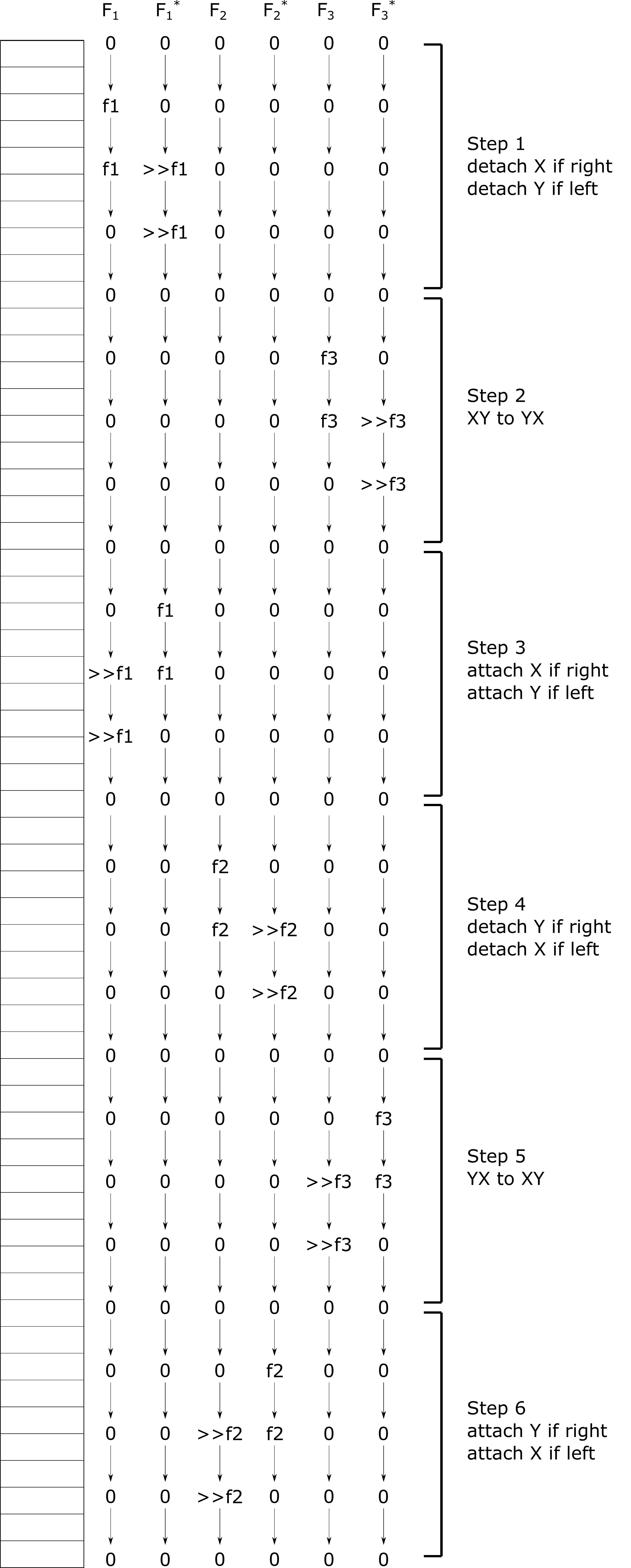}
	\caption{The protocol of fuel concentrations for the molecular motor that is inspired by actin myosin.}
	\label{fig:walkersteppingprotocol}
\end{figure}

This is more complicated than the previous set-up because the motor has more possible states but there are only 5 different reactions required. To move right there are the reactions
\begin{align}
	\fs[1]+\R+\x[\text{attached}]&\rlh\R+\x[\text{unattched}]+\f[1]\nn\\
	\fs[2]+\R+\y[\text{attached}]&\rlh\R+\y[\text{unattached}]+\f[2]\nn\\
	\fs[3]+\R+\x\y&\rlh\R+\y\x+\f[3]
\end{align}
where XY and YX represent the configurations where either the X foot is on the left of the motor or the Y foot is on the left of the motor. To move left the $\Le$ state catalyses the same equations but the fuels that cause the attaching and detaching of the X and Y are swapped and the same reaction to change the conformation of the walker is used
\begin{align}
	\fs[2]+\Le+\x[\text{attached}]&\rlh\Le+\x[\text{unattched}]+\f[2]\nn\\
	\fs[1]+\Le+\y[\text{attached}]&\rlh\Le+\y[\text{unattached}]+\f[1]
\end{align}
However, a more complicated protocol of fuel concentrations is required. This protocol is shown in figure \ref{fig:walkersteppingprotocol}. This protocol requires twice as many steps as the previous one.

\section{Examples}
\label{app:examples}
\subsection{2-state 3-symbol Machine}
This machine is claimed to be the smallest possible Turing machine \cite{smith2007universality} but it is only Turing complete if the tape initially has a complex encoding of an infinite string. The usual definition of a Turing machine requires that the tape is initialised with an infinitely repeating symbol containing only a finite string of different symbols \cite{neary2012complexity}. 

The molecules on the tape can be in three states: 0, 1 and 2. The `state molecule' of the head (like the `memory molecule') has two states: $\A$ and $\B$. The transitions for this machine are shown in table \ref{tab:wolfram}.
\begin{table}
\centering
\begin{tabular}{c | c c}
	& A & B \\ \hline
	0 & B1R & A2L \\
	1 & A2L & B2R \\
	2 & A1L & A0R
\end{tabular}
\caption{The rules for Wolfram's 2 state 3 symbol machine.}
\label{tab:wolfram}
\end{table}

As mentioned in the main text, the most straightforward way to convert this list of rules to an invertible map is for the memory molecule to be initially in the state $\m[0]$ and to convert it to a state labelled by the number of the rule that is used as below:
\begin{align}
	\A+\Le+0+\m[0]+\gs&\rlh\B+\R+1+\m[1]+\g\nonumber\\
	\A+\R+0+\m[0]+\gs&\rlh\B+\R+1+\m[2]+\g\nonumber\\
	\A+\Le+1+\m[0]+\gs&\rlh\A+\Le+2+\m[3]+\g\nonumber\\
	\A+\R+1+\m[0]+\gs&\rlh\A+\Le+2+\m[4]+\g\nonumber\\
	\A+\Le+2+\m[0]+\gs&\rlh\A+\Le+1+\m[5]+\g\nonumber\\
	\A+\R+2+\m[0]+\gs&\rlh\A+\Le+1+\m[6]+\g\nonumber\\
	\B+\Le+0+\m[0]+\gs&\rlh\A+\Le+2+\m[7]+\g\nonumber\\
	\B+\R+0+\m[0]+\gs&\rlh\A+\Le+2+\m[8]+\g\nonumber\\
	\B+\Le+1+\m[0]+\gs&\rlh\B+\R+2+\m[9]+\g\nonumber\\
	\B+\R+1+\m[0]+\gs&\rlh\B+\R+2+\m[10]+\g\nonumber\\
	\B+\Le+2+\m[0]+\gs&\rlh\A+\R+0+\m[11]+\g\nonumber\\
	\B+\R+2+\m[0]+\gs&\rlh\A+\R+0+\m[12]+\g.
	\label{eq:wolfram}
\end{align}
We only need one pair of fuel molecules because each reaction is otherwise unique. The protocol is to start with $[\g]=[\gs]=0$, then increase $[\g]$ up to some value then increase $[\gs]$ up to a value much greater than $[\g]$ then decrease $[\g]$ to zero, then decrease $[\gs]$ to zero.

There are no reactions that convert the directional molecule to the H state in equation \ref{eq:wolfram} because this machine does not have a halting state. This lack of a halting state is another way in which this machine differs from the standard definition of a Turing machine.

This scheme uses a state molecule with two states, a directional molecule with two states, a symbol molecule for each position on the main tape with three states, and a memory molecule at each position on the memory tape with 13 states. One pair of fuel molecule species is also needed (not including the stepping).

\begin{figure}
	\centering
	\includegraphics[width=0.5\linewidth]{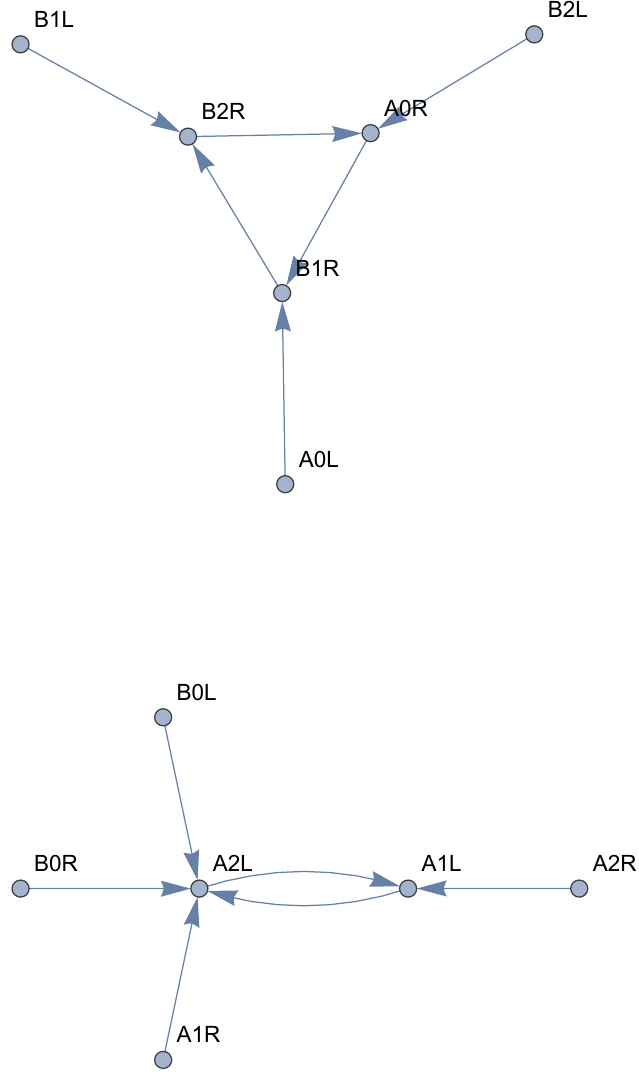}
	\caption{The transitions in a single step of the 2-state 3-symbol machine. The largest in degree is four.}
	\label{fig:wolframgraph}
\end{figure}

This naive scheme uses more memory states than necessary. Figure \ref{fig:wolframgraph} shows that the maximal in degree of any state is four so only $\m[1]$ to $\m[4]$ are needed to distinguish the previous state of the machine.


Therefore, we only need five states for the memory molecule and we can use the reactions:
\begin{align}
	\A+\Le+0+\m[0]+\gs&\rlh\B+\R+1+\m[1]+\g\nonumber\\
	\A+\R+0+\m[0]+\gs&\rlh\B+\R+1+\m[2]+\g\nonumber\\
	\A+\Le+1+\m[0]+\gs&\rlh\A+\Le+2+\m[1]+\g\nonumber\\
	\A+\R+1+\m[0]+\gs&\rlh\A+\Le+2+\m[2]+\g\nonumber\\
	\A+\Le+2+\m[0]+\gs&\rlh\A+\Le+1+\m[1]+\g\nonumber\\
	\A+\R+2+\m[0]+\gs&\rlh\A+\Le+1+\m[2]+\g\nonumber\\
	\B+\Le+0+\m[0]+\gs&\rlh\A+\Le+2+\m[3]+\g\nonumber\\
	\B+\R+0+\m[0]+\gs&\rlh\A+\Le+2+\m[4]+\g\nonumber\\
	\B+\Le+1+\m[0]+\gs&\rlh\B+\R+2+\m[1]+\g\nonumber\\
	\B+\R+1+\m[0]+\gs&\rlh\B+\R+2+\m[2]+\g\nonumber\\
	\B+\Le+2+\m[0]+\gs&\rlh\A+\R+0+\m[1]+\g\nonumber\\
	\B+\R+2+\m[0]+\gs&\rlh\A+\R+0+\m[2]+\g.
\end{align}

This scheme uses a state molecule with two states, a directional molecule with two states, a symbol molecule for each position on the main tape with three states, a memory molecule at each position on the memory tape with five states, and one pair of fuel molecule species.

Instead of a single molecule with five states we can use three molecules each with two states. Of course, this leaves multiple unused states. Use one molecule for direction of previous step ($\m[\Le]$/$\m[\R]$), one to distinguish the 2-to-1 rules ($\m[1]$/$\m[2]$) and one to give reaction direction/take place of hidden states ($\m[i]$/$\m[f]$). We are free to choose the initial state of the ($\m[\Le]$/$\m[\R]$) and ($\m[1]$/$\m[2]$) molecules so we arbitrarily choose $\m[\Le]$ and $\m[1]$.

Therefore, reactions become:
\begin{align}
	\A+\Le+0+\m[i]+\m[\Le]+\m[1]+\gs&\rlh\B+\R+1+\m[f]+\m[\Le]+\m[1]+\g\nonumber\\
	\A+\R+0+\m[i]+\m[\Le]+\m[1]+\gs&\rlh\B+\R+1+\m[f]+\m[\R]+\m[1]+\g\nonumber\\
	\A+\Le+1+\m[i]+\m[\Le]+\m[1]+\gs&\rlh\A+\Le+2+\m[f]+\m[\Le]+\m[1]+\g\nonumber\\
	\A+\R+1+\m[i]+\m[\Le]+\m[1]+\gs&\rlh\A+\Le+2+\m[f]+\m[\R]+\m[1]+\g\nonumber\\
	\A+\Le+2+\m[i]+\m[\Le]+\m[1]+\gs&\rlh\A+\Le+1+\m[f]+\m[\Le]+\m[1]+\g\nonumber\\
	\A+\R+2+\m[i]+\m[\Le]+\m[1]+\gs&\rlh\A+\Le+1+\m[f]+\m[\R]+\m[1]+\g\nonumber\\
	\B+\Le+0+\m[i]+\m[\Le]+\m[1]+\gs&\rlh\A+\Le+2+\m[f]+\m[\Le]+\m[2]+\g\nonumber\\
	\B+\R+0+\m[i]+\m[\Le]+\m[1]+\gs&\rlh\A+\Le+2+\m[f]+\m[\R]+\m[2]+\g\nonumber\\
	\B+\Le+1+\m[i]+\m[\Le]+\m[1]+\gs&\rlh\B+\R+2+\m[f]+\m[\Le]+\m[1]+\g\nonumber\\
	\B+\R+1+\m[i]+\m[\Le]+\m[1]+\gs&\rlh\B+\R+2+\m[f]+\m[\R]+\m[1]+\g\nonumber\\
	\B+\Le+2+\m[i]+\m[\Le]+\m[1]+\gs&\rlh\A+\R+0+\m[f]+\m[\Le]+\m[1]+\g\nonumber\\
	\B+\R+2+\m[i]+\m[\Le]+\m[1]+\gs&\rlh\A+\R+0+\m[f]+\m[\R]+\m[1]+\g
\end{align}
Clearly, our trade-off for reducing the number of states is to have reactions of six molecules.

The situation can be improved because we do not need all of the molecules in all of the reactions. We only need the $\m[1]$/$\m[2]$ reaction in the rule that needs it. Therefore, the reactions are:
\begin{align}
	\A+\Le+0+\m[i]+\m[\Le]+\gs&\rlh\B+\R+1+\m[f]+\m[\Le]+\g\nonumber\\
	\A+\R+0+\m[i]+\m[\Le]+\gs&\rlh\B+\R+1+\m[f]+\m[\R]+\g\nonumber\\
	\A+\Le+1+\m[i]+\m[\Le]+\m[1]+\gs&\rlh\A+\Le+2+\m[f]+\m[\Le]+\m[1]+\g\nonumber\\
	\A+\R+1+\m[i]+\m[\Le]+\m[1]+\gs&\rlh\A+\Le+2+\m[f]+\m[\R]+\m[1]+\g\nonumber\\
	\A+\Le+2+\m[i]+\m[\Le]+\gs&\rlh\A+\Le+1+\m[f]+\m[\Le]+\g\nonumber\\
	\A+\R+2+\m[i]+\m[\Le]+\gs&\rlh\A+\Le+1+\m[f]+\m[\R]+\g\nonumber\\
	\B+\Le+0+\m[i]+\m[\Le]+\m[1]+\gs&\rlh\A+\Le+2+\m[f]+\m[\Le]+\m[2]+\g\nonumber\\
	\B+\R+0+\m[i]+\m[\Le]+\m[1]+\gs&\rlh\A+\Le+2+\m[f]+\m[\R]+\m[2]+\g\nonumber\\
	\B+\Le+1+\m[i]+\m[\Le]+\gs&\rlh\B+\R+2+\m[f]+\m[\Le]+\g\nonumber\\
	\B+\R+1+\m[i]+\m[\Le]+\gs&\rlh\B+\R+2+\m[f]+\m[\R]+\g\nonumber\\
	\B+\Le+2+\m[i]+\m[\Le]+\gs&\rlh\A+\R+0+\m[f]+\m[\Le]+\g\nonumber\\
	\B+\R+2+\m[i]+\m[\Le]+\gs&\rlh\A+\R+0+\m[f]+\m[\R]+\g.
\end{align}

\subsection{4-state 6-symbol Machine}
\label{ap:Rogozhin}
There is a trade-off between the number of head states and the number of tape symbols \cite{neary2012complexity}. The machine with the smallest number of different transition rules, which correspond to chemical reactions, is the four state six symbol machine from Rogozhin \cite{rogozhin1996small}. The transitions for this machine are shown in table \ref{tab:Rogozhin}. A is the initial head state and 4 is the blank symbol.

\begin{table}[h]
\centering
\begin{tabular}{c | c c c c}
	& A & B & C & D \\ \hline
	0 & A3L & B4R & C0R & D4R \\
	1 & A2R & C2L & D3R & B5L \\
	2 & A1L & B3R & C1R & D3R \\
	3 & A4R & B2L & H   & H   \\
	4 & A3L & B0L & A5R & B5L \\
	5 & D4R & B1R & A0R & D1R
\end{tabular}
\caption{The rules for Rogozhin's universal 4 state 6 symbol machine (with relabelled states and symbols).The letters are the states and the numbers are the symbols.}
\label{tab:Rogozhin}
\end{table}

\begin{figure}
	\centering
	\includegraphics[width=\linewidth]{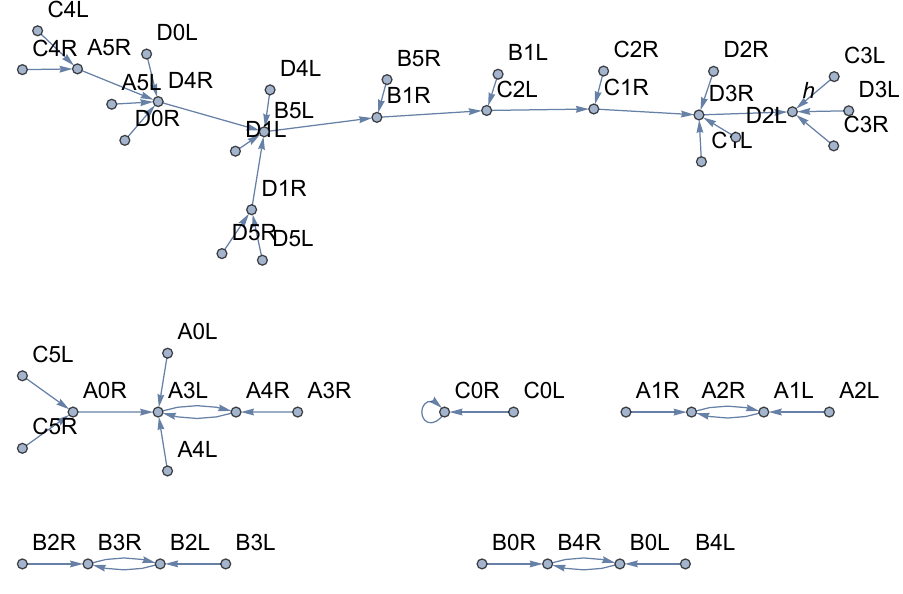}
	\caption{Graph showing one step of Rogozhin's 4-state 6-symbol machine. There are 5 nodes with an in degree of 4.}
	\label{fig:rogozhingraph}
\end{figure}

We see in figure \ref{fig:rogozhingraph} that the largest in degree is 4 so only 5 memory states are required. The reactions are:
\clearpage
\begin{align}
		\A+0+\Le+\m[0]+\m[1]+\gs&\rlh\A+3+L+\m[1]+\g \nn\\
		\A+0+\R+\m[0]+\m[2]+\gs&\rlh\A+3+L+\m[2]+\g \nn\\
		\B+0+\Le+\m[0]+\m[1]+\gs&\rlh\B+4+R+\m[1]+\g \nn\\
		\B+0+\R+\m[0]+\m[2]+\gs&\rlh\B+4+R+\m[2]+\g \nn\\
		\C+0+\Le+\m[0]+\m[1]+\gs&\rlh\C+0+R+\m[1]+\g \nn\\
		\C+0+\R+\m[0]+\m[2]+\gs&\rlh\C+0+R+\m[2]+\g \nn\\
		\D+0+\Le+\m[0]+\m[1]+\gs&\rlh\D+4+R+\m[1]+\g \nn\\
		\D+0+\R+\m[0]+\m[2]+\gs&\rlh\D+4+R+\m[2]+\g \nn\\
		\A+1+\Le+\m[0]+\m[1]+\gs&\rlh\A+2+R+\m[1]+\g \nn\\
		\A+1+\R+\m[0]+\m[2]+\gs&\rlh\A+2+R+\m[2]+\g \nn\\
		\B+1+\Le+\m[0]+\m[1]+\gs&\rlh\C+2+L+\m[1]+\g \nn\\
		\B+1+\R+\m[0]+\m[2]+\gs&\rlh\C+2+L+\m[2]+\g \nn\\
		\C+1+\Le+\m[0]+\m[1]+\gs&\rlh\D+3+R+\m[1]+\g \nn\\
		\C+1+\R+\m[0]+\m[2]+\gs&\rlh\D+3+R+\m[2]+\g \nn\\
		\D+1+\Le+\m[0]+\m[1]+\gs&\rlh\B+5+L+\m[1]+\g \nn\\
		\D+1+\R+\m[0]+\m[2]+\gs&\rlh\B+5+L+\m[2]+\g \nn\\
		\A+2+\Le+\m[0]+\m[1]+\gs&\rlh\A+1+L+\m[1]+\g \nn\\
		\A+2+\R+\m[0]+\m[2]+\gs&\rlh\A+1+L+\m[2]+\g \nn\\
		\B+2+\Le+\m[0]+\m[1]+\gs&\rlh\B+3+R+\m[1]+\g \nn\\
		\B+2+\R+\m[0]+\m[2]+\gs&\rlh\B+3+R+\m[2]+\g \nn\\
		\C+2+\Le+\m[0]+\m[1]+\gs&\rlh\C+1+R+\m[1]+\g \nn\\
		\C+2+\R+\m[0]+\m[2]+\gs&\rlh\C+1+R+\m[2]+\g \nn\\
		\D+2+\Le+\m[0]+\m[1]+\gs&\rlh\D+3+R+\m[1]+\g \nn\\
		\D+2+\R+\m[0]+\m[2]+\gs&\rlh\D+3+R+\m[2]+\g \nn\\
		\A+3+\Le+\m[0]+\m[1]+\gs&\rlh\A+4+R+\m[1]+\g \nn\\
		\A+3+\R+\m[0]+\m[2]+\gs&\rlh\A+4+R+\m[2]+\g \nn\\
		\B+3+\Le+\m[0]+\m[1]+\gs&\rlh\B+2+L+\m[1]+\g \nn\\
		\B+3+\R+\m[0]+\m[2]+\gs&\rlh\B+2+L+\m[2]+\g \nn\\
		\C+3+\Le+\m[0]+\m[1]+\gs&\rlh\C+3+H+\m[1]+\g \nn\\
		\C+3+\R+\m[0]+\m[2]+\gs&\rlh\C+3+H+\m[2]+\g \nn\\
		\D+3+\Le+\m[0]+\m[1]+\gs&\rlh\D+3+H+\m[1]+\g \nn\\
		\D+3+\R+\m[0]+\m[2]+\gs&\rlh\D+3+H+\m[2]+\g \nn\\
		\A+4+\Le+\m[0]+\m[1]+\gs&\rlh\A+3+L+\m[1]+\g \nn\\
		\A+4+\R+\m[0]+\m[2]+\gs&\rlh\A+3+L+\m[2]+\g \nn\\
		\B+4+\Le+\m[0]+\m[1]+\gs&\rlh\B+0+L+\m[1]+\g \nn\\
		\B+4+\R+\m[0]+\m[2]+\gs&\rlh\B+0+L+\m[2]+\g \nn\\
		\C+4+\Le+\m[0]+\m[1]+\gs&\rlh\A+5+R+\m[1]+\g \nn\\
		\C+4+\R+\m[0]+\m[2]+\gs&\rlh\A+5+R+\m[2]+\g \nn\\
		\D+4+\Le+\m[0]+\m[1]+\gs&\rlh\B+5+L+\m[1]+\g \nn\\
		\D+4+\R+\m[0]+\m[2]+\gs&\rlh\B+5+L+\m[2]+\g \nn\\
		\A+5+\Le+\m[0]+\m[1]+\gs&\rlh\D+4+R+\m[1]+\g \nn\\
		\A+5+\R+\m[0]+\m[2]+\gs&\rlh\D+4+R+\m[2]+\g \nn\\
		\B+5+\Le+\m[0]+\m[1]+\gs&\rlh\B+1+R+\m[1]+\g \nn\\
		\B+5+\R+\m[0]+\m[2]+\gs&\rlh\B+1+R+\m[2]+\g \nn\\
		\C+5+\Le+\m[0]+\m[1]+\gs&\rlh\A+0+R+\m[1]+\g \nn\\
		\C+5+\R+\m[0]+\m[2]+\gs&\rlh\A+0+R+\m[2]+\g \nn\\
		\D+5+\Le+\m[0]+\m[1]+\gs&\rlh\D+1+R+\m[1]+\g \nn\\
		\D+5+\R+\m[0]+\m[2]+\gs&\rlh\D+1+R+\m[2]+\g \nn\\
\end{align}

\section{Cost per time complexity}
\label{ap:cost_per_time}
Instead of counting the number of different computations done we could instead measure the total amount of time spent doing the computations. This is how long all of the computations would take if run in series on a single machine. By `time' we mean not the physical time but the number of steps the Turing machines take.

In the set of computations of interest \(\mathcal{S}\) each computation \(i\) has a time to halt of \(t_i\). As in the main text all \(M\) computations in set \(\mathcal{S}\) halt within a time of \(t_M\) so \(t_i<t_M\) for the \(i\) in this set.

As in the main text the cost is $C(M,t) = \alpha(M) t + C_0(M)$, with $C_0$ representing initiation and termination costs.

The amount of time spent doing the computations in the set \(\mathcal{S}_M\) is simply the sum of the times to halt of each of the computations
\begin{equation}
    A(M) = \sum_{\mathcal{S}_M} t_i.
\end{equation}
i.e. this the time multiplied by the number of machines but not counting the time each machine spends in the halt state.

So now the quantity to be considered is the cost per amount of time spent doing the computations \(C(M,t)/A(M)\). This is the cost per useful step of a machine. We want to know if this quantity can be made arbitrarily small by increasing the number of computations.

The computations take at least one step of the Turing machine so the total amount of time spent computing is lower bounded by the number of machines, \(A(M) \geq M\), so the upper bound on the cost per amount of time spend doing the computation is the cost per computation, \( C(M,t)/A(M) \leq C(M,t)/M \). Therefore, the constraint on the set of computations is easier to satisfy than he constraint in the main text that \(\mathcal{S}_M\) must be such that \(t_M\) grows more slowly than \(M^{0.5}\).

The constraint in this case is that \(\frac{\sum_{\mathcal{S}_M} t_i}{t_M}\) must grow more slowly than \(M^{0.5}\).

\end{document}